\documentclass[aps,prl,amsmath,amssymb,reprint,superscriptaddress,showpacs]{revtex4-1}

\usepackage{graphicx}
\usepackage{dcolumn}
\usepackage{bm}
\usepackage{hyperref}
\usepackage{color}
\usepackage{soul}

\begin{document}


\title{Quantum criticality in $A$Fe$_2$As$_2$ with $A$~= K, Rb, and Cs \\suppresses superconductivity}


\author{Felix Eilers}
\author{Kai Grube}
\author{Diego A. Zocco}
\author{Thomas Wolf}
\author{Michael Merz}
\author{Peter Schweiss}
\author{Rolf Heid}
\author{Robert Eder}
\affiliation{Institut f\"ur Festk\"orperphysik, Karlsruher Institut f\"ur Technologie, 76021 Karlsruhe, Germany}
\author{Rong Yu}
\affiliation{Department of Physics, Renmin University of China, Beijing 100872, China}
\author{Jian-Xin Zhu}
\affiliation{Theoretical Division, Los Alamos National Laboratory, Los Alamos, New Mexico 87545, USA}
\author{Qimiao Si}
\affiliation{Department of Physics and Astronomy, Rice University, Houston, TX 77005, USA}
\author{Takasada Shibauchi}
\affiliation{Department of Advanced Materials Science, University of Tokyo, Kashiwa, Chiba 277-8561, Japan}
\affiliation{Department of Physics, Kyoto University, Sakyo-ku, Kyoto 606-8502, Japan}
\author{Hilbert v. L\"ohneysen}
\affiliation{Institut f\"ur Festk\"orperphysik, Karlsruher Institut f\"ur Technologie, 76021 Karlsruhe, Germany}
\affiliation{Physikalisches Institut, Karlsruhe Institut f\"ur Technologie, 76049 Karlsruhe, Germany}


\date{\today}

\begin{abstract}
Superconductors close to quantum phase transitions often exhibit a simultaneous increase of electronic correlations and superconducting transition temperatures. Typical examples are given by the recently discovered iron-based superconductors. We investigated the band-specific quasiparticle masses of $A$Fe$_2$As$_2$ single crystals with $A$~=~K, Rb, and Cs and determined their pressure dependence. The evolution of electronic correlations could be tracked as a function of volume and hole doping. The results indicate that with increasing alkali-metal ion radius a quantum critical point is approached. The critical fluctuations responsible for the enhancement of the quasiparticle masses appear to suppress the superconductivity. 
\end{abstract}

 

\maketitle

Unconventional superconductivity (SC) often emerges in the proximity of continuous, zero-temperature phase transitions, so-called quantum critical points (QCPs). In particular, the onset of magnetic order is generally believed to drive SC by magnetic quantum criticality. Examples encompass the cuprates, organic metals, heavy-fermion systems, and the recently discovered iron-based superconductors. A particularly illustrative example is given by BaFe$_2$(As$_{1-x}$P$_x$)$_2$. Here, the application of chemical pressure, by replacing As with isovalent, smaller P ions, suppresses antiferromagnetic (AF) order resulting in an extended superconducting dome with a maximum transition temperature $T_c \approx 30\,$K at the critical concentration $x_c = 0.3$ \cite{Shibauchi2014}. The QCP at $x_c$ shielded by SC was anticipated theoretically \cite{Dai2009} and observed through strongly enhanced quasiparticle masses and deviations from Fermi-liquid (FL) behavior. 
In this Letter we show that the isostructural superconductor KFe$_2$As$_2$ can likewise be pushed towards a QCP by substituting isovalent Rb and Cs for K. In these compounds, in contrast to the examples listed above and despite general consensus, the proximity to a QCP appears to suppress SC.

\begin{figure}
\includegraphics{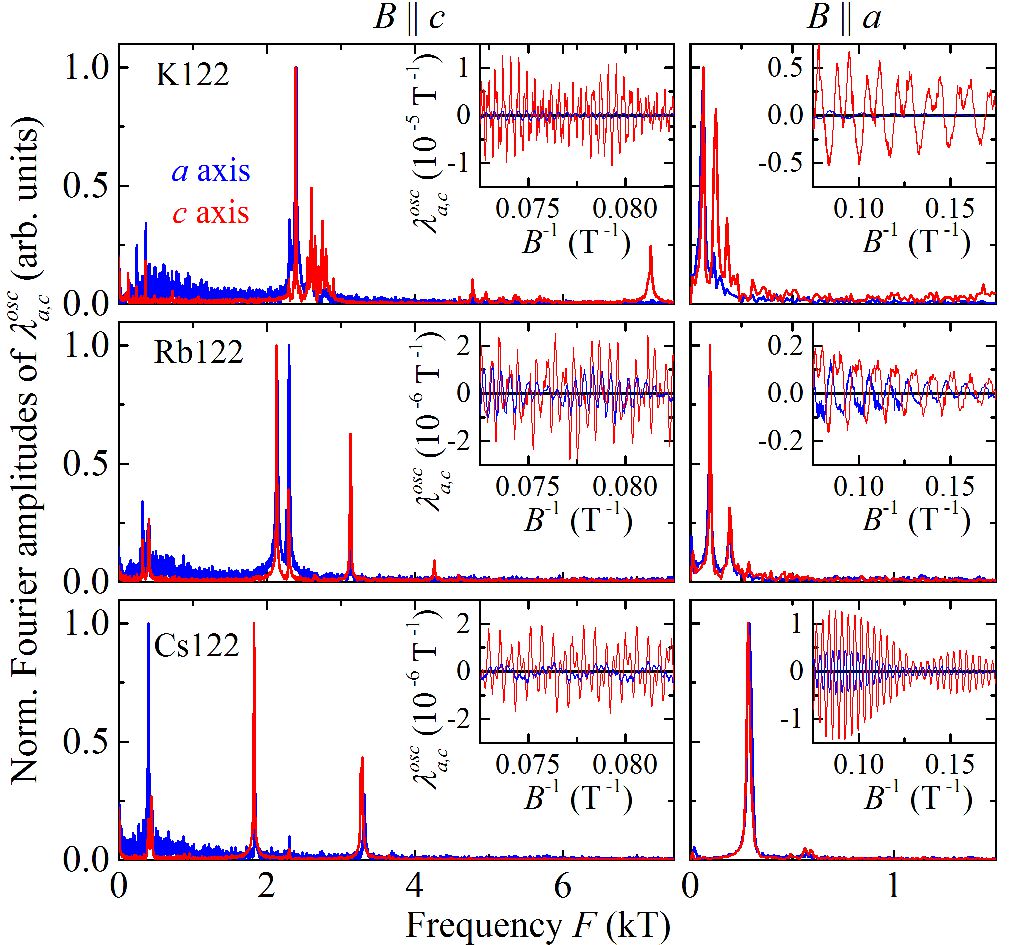}%
\caption{\label{fig:FFT} (Color online) Normalized Fourier spectra of the oscillatory part of the magnetostriction coefficient $\lambda^{osc}_{i}$ (insets) along the $i=a$ (blue) and $c$ axis (red line) of $A$122, $A$~= K, Rb, and Cs, at $T=50$\,mK for $B \parallel c$ and $a$.}
\end{figure}

The alkali metal series $A$Fe$_2$As$_2$ ($A$122) with $A$~= K, Rb, and Cs represents one of the rare examples of stoichiometric iron-arsenide superconductors. According to LDA calculations their low $T_c$ values of less than $3.5\,$K cannot be explained by electron-phonon coupling. Angle-resolved photoemission spectroscopy and thermal conductivity measurements suggest an unconventional pairing mechanism \cite{Okazaki2012,Reid2012,Hong2013,Zhang2014}. Recent specific-heat measurements reveal huge Sommerfeld coefficients $\gamma$ which exceed those of BaFe$_2$(As$_{1-x}$P$_x$)$_2$ in apparent contradiction to the low $T_c$ values \cite{Hardy2013,Zhang2014,Wang2013}. 
In order to elucidate the highly correlated normal state and its relationship to SC, we investigated the quantum oscillations observable in the magnetostriction of the $A$122 series (insets of Fig.~\ref{fig:FFT}).

Single crystalline samples were grown from an arsenic-rich flux yielding $T_c$ values of 3.4\,K, 2.5\,K, and 2.25\,K for $A$~= K, Rb, and Cs, respectively \cite{Suppl}. Their crystal structures were analyzed with four-circle X-ray diffraction at room temperature. Structural refinement confirmed the space group to be $I4/mmm$, type ThCr$_2$Si$_2$, and the composition to be stoichiometric within the experimental error of 1-2\,\%.
The low-impurity concentration of the crystals allows us to observe quantum oscillations of the sample length as a function of the applied magnetic field $B=\mu_0H$ ranging between the upper critical field $B_{c2}$ and the maximum field of 14\,T. To extract the fundamental frequencies we performed a Fourier transformation of the oscillatory part of the measured magnetostriction coefficient $\lambda_i \equiv L_i^{-1} \partial L_i / \partial B$, where $L_i$ is the sample length along the crystallographic $i=a$ and $c$ directions. 

\begin{figure}
\includegraphics{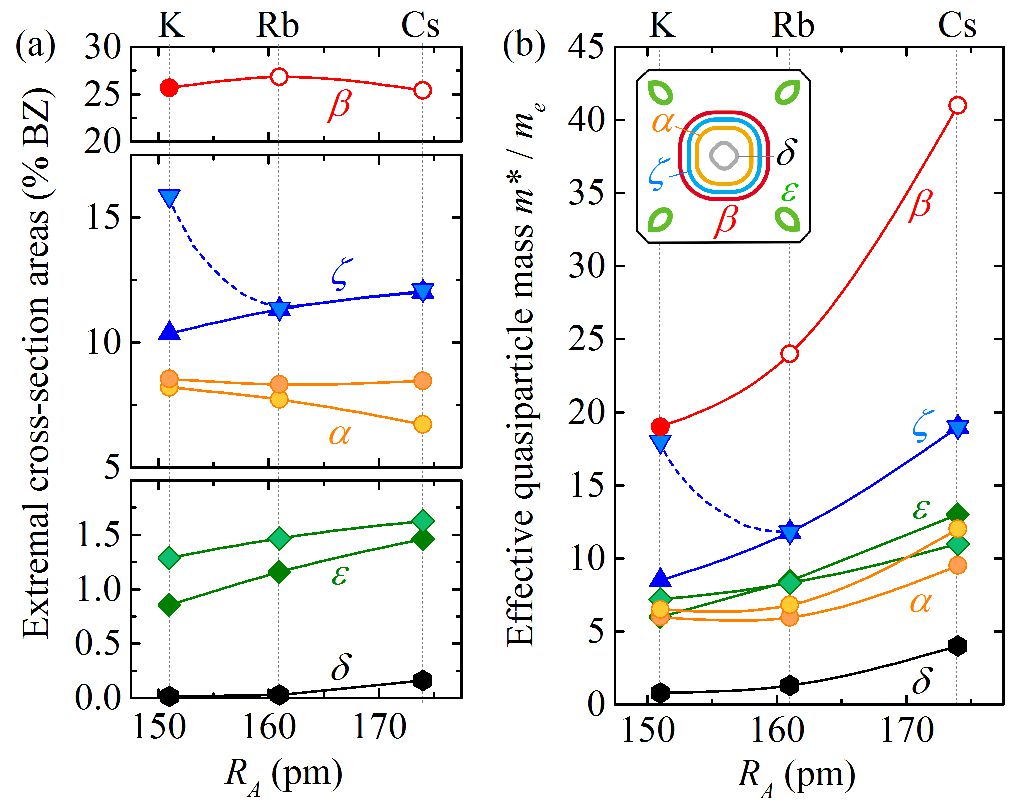}%
\caption{\label{fig:mass} (Color online) (a) Extremal cross-sectional areas of the Fermi surfaces of (K,Rb,Cs)122, expressed as fractions of the first Brillouin-zone volumes. (b) Effective quasiparticle masses $m^*$, plotted against the ionic radius of the alkaline atom $R_A$. The values presented by open symbols were obtained by assuming a constant total hole count and using the Sommerfeld coefficients \cite{Hardy2013,Zhang2014,Wang2013}. Lines are guides to the eye.}
\end{figure}

The spectra at $T=50\,$mK for both field directions, $B \parallel a$ and $c$, are plotted as amplitude against frequency $F$ in Fig.\,\ref{fig:FFT}. The measurements of K122 were taken from the work previously published \cite{Zocco2013,Zocco2014}. The difference between the spectra along $a$ and $c$ axes reflects the anisotropic uniaxial pressure dependences of the Fermi-surface (FS) cross-sections \cite{Shoenberg1984}. The observed sharp peaks mark the fundamental and higher harmonic cyclotron frequencies corresponding to the extremal orbits of the FS, which have been assigned to specific bands by comparing the spectra of Rb122 and Cs122 to those of K122 and to our LDA calculations. For (Rb,Cs)122, all the FS sheets could be identified apart from the $\beta$ band, which has the largest  cross-section area. As the isovalent substitution of $A$ keeps the total hole count constant we can estimate the contribution of the $\beta$ sheet by subtracting the cross-sections of all other bands from the FS volume of K122. In Fig.~\ref{fig:mass}(a) the obtained extremal cross-sectional areas, expressed as fractions of the volume of the first Brillouin zone, are plotted against the eight-fold coordinated ionic radius of the alkali atom $R_A$. 

In line with our LDA calculations, the data show no change of the FS topology with increasing $R_A$. The two-dimensionality of the FeAs-layered structure is reflected by three tubes in the Brillouin-zone center [$\alpha$, $\zeta$, and $\beta$, see inset of Fig.\,\ref{fig:mass}(b)] and one tube in each corner ($\epsilon$). Marginal three-dimensional features were inferred from closely spaced peaks in the Fourier spectra, indicating a gentle warping of the FS tubes, and from a single fundamental frequency for $B \parallel a$, attributed to a tiny pocket at the top center. So far, this pocket has only been seen in magnetostriction measurements which are especially sensitive to small extremal FS areas \cite{Zocco2014}.

The decay of the oscillations with increasing $T$ \cite{Suppl} is used to determine the effective quasiparticle masses $m_j^*$ of each FS sheet $j$. Since in quasi-two-dimensional systems the Sommerfeld coefficient is given by $\gamma \approx (\pi k_{\text{B}}^2N_{\text{A}}a^2/3\hbar^2) \sum_j{m_j^*}$ ($N_{\text{A}}$: Avogadro's number), the mass of the $\beta$ tube can be determined by subtracting the contributions of all other bands from the published $\gamma$ values. The obtained large $m^*_{\beta}$ values suggest that the base temperature of $15$\,mK of our experiment was too high to allow the observation of the corresponding frequencies. Fig.~\ref{fig:mass}(b) summarizes the resulting quasiparticle masses as a function of $R_A$. 
Not only $m^*_{\beta}$, but all effective masses exhibit significant increases with $R_A$, with a factor of $m^*_j(\text{Cs})/m^*_j(\text{K})\approx 2$ which interestingly is similar to that of $\gamma(\text{Cs})/\gamma(\text{K})$. 
LDA calculations, on the other hand, predict band masses an order of magnitude smaller, with a comparatively slight increase towards Cs122. This difference reinforces the assumption that the huge, low-temperature specific heat arises from strong electronic correlations (see below).

\begin{figure}[t!]
\includegraphics[width=8.6cm]{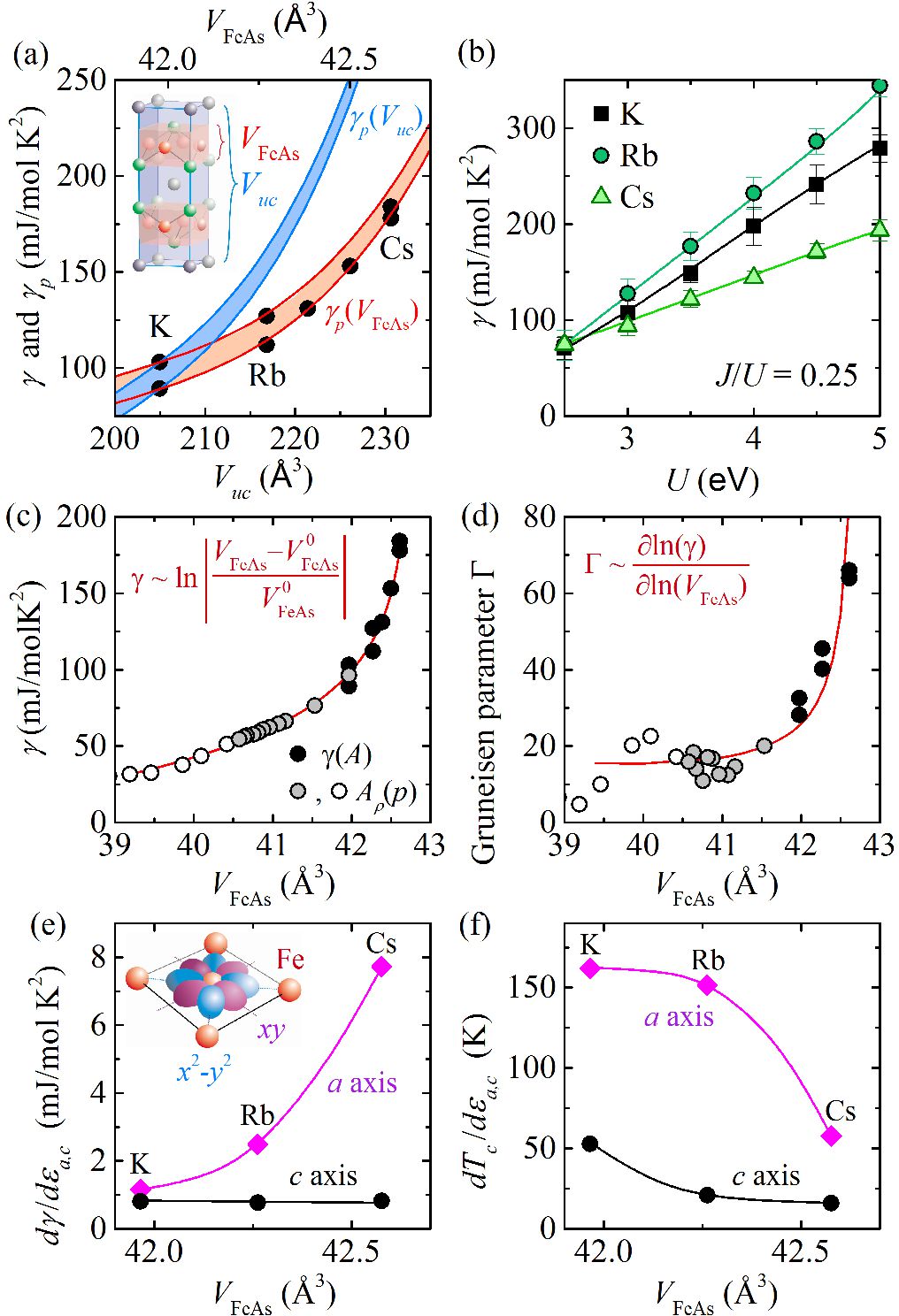}%
\caption{\label{fig:all} (Color online) (a) Comparison between chemical (black dots) and hydrostatic pressure dependences of the Sommerfeld coefficient $\gamma$ as a function of the unit-cell volume $V_{uc}$ and that of a single FeAs layer $V_{\text{FeAs}}$. The hydrostatic pressure dependences have been extracted from the volume thermal expansion $\alpha_V$ for $V=V_{uc}$ (blue lines) and $V_{\text{FeAs}}$ (red lines). (b) The calculated Sommerfeld coefficient $\gamma$ as a function of $U$ in a multiorbital Hubbard model for $A$122 ($A$~= K, Rb, Cs) (see text). (c) Evolution of $\gamma$ with increasing $V_{\text{FeAs}}$. The grey \cite{Taufour2014} and open dots \cite{Ying2015} are extracted from high-pressure resistivity measurements. The red line is a fit to the data points. (d) The Gr\"uneisen parameter $\Gamma$, calculated from $\gamma$, $A_{\rho}$ and $\alpha_V$. The red line is determined with the fit shown in (d) under the assumption of a Fermi liquid. (e) Strain dependences of $\gamma$. The inset shows the Fe plane with the $3d_{xy}$ and $3d_{x^2-y^2}$ orbitals. (f) Strain dependences of the superconducting transition temperature $T_c$.
}
\end{figure}

The negative chemical pressure exerted by replacing K with the larger Rb or Cs ions mainly expands the lattice constant $c$ by pushing the FeAs layers further apart from each other. 
To elucidate the difference between chemical and external, hydrostatic pressure, we performed thermal-expansion measurements. 
In a FL, the uniaxial pressure dependence of the Sommerfeld coefficient is related to the linear thermal-expansion coefficient $\alpha_i$ by $\partial \gamma/\partial p_i = -(V/T)\alpha_i$, where $V$ is the molar volume and $p_i$ uniaxial pressure in $i=a,c$ direction. The hydrostatic pressure dependence is given by $\partial \gamma/\partial p = 2\partial \gamma/\partial p_a + \partial \gamma/\partial p_c$. 
We found that, between $T_c$ and 4\,K, all three compounds show FL behavior, i.e., constant $\alpha_i/T$ values, similarly to the published $C/T$ data. With increasing $R_A$ and, consequently, growing unit-cell volume, the pressure dependence decreases from $\partial \gamma/\partial p = -20.3$\,mJ/molK$^2$GPa (K)\cite{Burger2013} to $-36.6$\,mJ/molK$^2$GPa (Rb) and $-77.7$\,mJ/molK$^2$GPa (Cs) suggesting a nonlinear $V$ dependence of $\gamma$.
In order to compare with the chemical pressure effect, we convert the hydrostatic pressure dependences from our thermal expansion measurements to volume derivatives $\partial \gamma/\partial V = -VB_T\partial \gamma/\partial p$, and estimate $\gamma_p(V)$ by integrating over the obtained $\partial\gamma(V)/\partial V$ values. Here, $B_T=-V\partial p/\partial V$ is the bulk modulus. The integration constant is provided by the published Sommerfeld coefficients of K122. 

In a first step we take $V$ to be the unit-cell volume $V_{uc}$ using $B_T(V_{uc})\approx 40$\,GPa of K122 \cite{Tafti2014}. 
The calculated $\gamma_p(V_{uc})$ curves are displayed in Fig.\,\ref{fig:all}(a) together with the Sommerfeld coefficients $\gamma(V_{uc})$ at ambient pressure \cite{Hardy2013,Kanter2010,Zhang2014,Wang2013}. Obviously, $\gamma_p(V_{uc})$ strongly overestimates the chemical pressure effect $\gamma(V_{uc})$. In this simple estimate we have neglected that chemical and hydrostatic pressures affect the crystal structure in different ways: while the latter leads to a reduction or increase of \textit{all} bond lengths, the alkali-metal substitution changes the $c$ length and FeAs-layer thickness $h_{\text{FeAs}}$ with opposite trends: it increases $c$ and decreases $h_{\text{FeAs}}$ \cite{Suppl}. 
Therefore, in a second step, we perform the comparison on the basis of the FeAs-cell volume $V_{\text{FeAs}}=a^2h_{\text{FeAs}}$. The related bulk modulus of $B_T(V_{\text{FeAs}}) \approx 145$\,GPa was inferred from the high-pressure data of K122 from Ref.\,\cite{Tafti2014}. The estimated $\gamma_p(V_{\text{FeAs}})$ curves are likewise shown in Fig.\,\ref{fig:all}(a). Taking $V_{\text{FeAs}}$ as the decisive pressure-dependent volume, hydrostatic and chemical pressure dependences coincide with each other. This agreement suggests that the enhanced correlations originate from a change of the direct Fe environment, most probably due to a reduced hybridization of the Fe 3$d$ states with nearest-neighbor Fe or As orbitals.

The important role of the FeAs-cell volume provides the basis to study the evolution of electronic correlations in a wider phase space. We extended the pressure dependence of $\gamma$ to smaller $V_{\text{FeAs}}$ values by resorting to resistivity measurements of K122 under hydrostatic pressure \cite{Taufour2014,Ying2015}. Since the FL state of K122 follows the Kadowaki-Woods (KW) relation \cite{Hashimoto2010} we relate $\gamma$ to the scattering cross section $A_{\rho}$ of the low-temperature resistivity $\rho = \rho_0+A_{\rho} T^2$ by using a proportionality factor of $A_{\rho}/\gamma^2 \approx 2\cdot 10^{-6}\,\mu\Omega$cm(Kmol/mJ)$^2$ so that the ambient pressure measurements are reproduced \cite{Hashimoto2010}. The extended $\gamma(V_{\text{FeAs}})$ data displayed in Fig.\,\ref{fig:all}(c) exhibit the typical sudden increase of a system that is tuned towards a QCP. The best fit to this mass divergence is given by a logarithmic volume dependence, $\gamma\propto \ln(\left| V_{\text{FeAs}}-V^0_{\text{FeAs}}\right|/V^0_{\text{FeAs}})$ with $V^0_{\text{FeAs}}=42.72\,$\AA$^3$, as proposed for a two-dimensional FL close to a Mott transition and found for BaFe$_2$(As$_{1-x}$P$_x$)$_2$ \cite{Abrahams2011}. 

QCPs are characterized by a vanishing characteristic energy scale $E^*$, which leads to a divergence of the Gr\"uneisen parameter $\Gamma \approx -d\ln(E^*)/\ln(V)$ for a pressure-induced QCP. We calculated $\Gamma = V_{\text{FeAs}}B_T(V_{\text{FeAs}})\alpha_V/\gamma$ from our thermal expansion data $\alpha_V=2\alpha_a+\alpha_c$. For the high-pressure resistivity measurements we use $\Gamma=d\ln(\gamma)/d\ln(V_{\text{FeAs}})=(1/2)d\ln(A_{\rho})/d\ln(V_{\text{FeAs}})$ by virtue of the KW relation. The volume dependence of $\Gamma$, shown in Fig.\,\ref{fig:all}(d), clearly exhibits a pronounced divergence. This provides clear evidence for the proximity of Cs122 to a $p$-induced QCP.

To specify the hybridized orbitals responsible for the critical mass enhancement, we determine the uniaxial pressure dependences $\partial\gamma/\partial p_i$ for $i=a,c$ from $\alpha_i$. Furthermore, we approximate the elastic constants $c_{ij}$ of the $A$122 compounds by DFT calculations \cite{Suppl} to obtain the strain dependences $d\gamma/d\epsilon_i=\sum c_{ij}d\gamma/dp_j$. The $d\gamma/d\epsilon_i$ values show that with increasing $V_{\text{FeAs}}$, mainly $a$-axis changes account for the mass enhancement [Fig.\,\ref{fig:all}(e)]. In particular, the divergence can only be observed in $d\gamma/d\epsilon_a$.
This allows identifying the orbitals involved, as in the FeAs-cell volume, only the Fe $3d_{xy}$ or $3d_{x^2-y^2}$ states are confined to the $ab$ plane [inset of Fig.\,\ref{fig:all}(e)] and, therefore, affected by $a$ or $b$-axis changes. 
Since the band-specific masses shown in Fig.\,\ref{fig:mass} reveal the heaviest masses for the $\beta$ bands with dominating $d_{xy}$ character, we conclude that the critical mass enhancement can be attributed to the hybridization of the in-plane $d_{xy}$ orbitals.

If the SC is supported by critical fluctuations arising close to a QCP, the strain dependence of the superconducting transition temperature $\partial T_c/\partial \epsilon_i$ should follow that of $\partial \gamma/\partial \epsilon_i$. To check this scenario, we determine the uniaxial pressure dependences of $T_c$ by using the Ehrenfest relations with the discontinuities of $\alpha_i$ and $C$ at $T_c$, and convert them to $\partial T_c/\partial \epsilon_i$. Similarly to $\partial \gamma/\partial \epsilon_c$, changes of the $c$ axis hardly affect $T_c$ [Fig.\,\ref{fig:all}(f)]. Contrary to the expectation of the above scenario, however, $\partial T_c/\partial \epsilon_a$ shows a behavior opposite to that of $\partial \gamma/\partial \epsilon_a$, and decreases with increasing $V_{\text{FeAs}}$. Apparently, the $d_{xy}$ states that drive the mass enhancement tend to simultaneously suppress the superconducting state. 

\begin{figure}
\includegraphics[width=8.6cm]{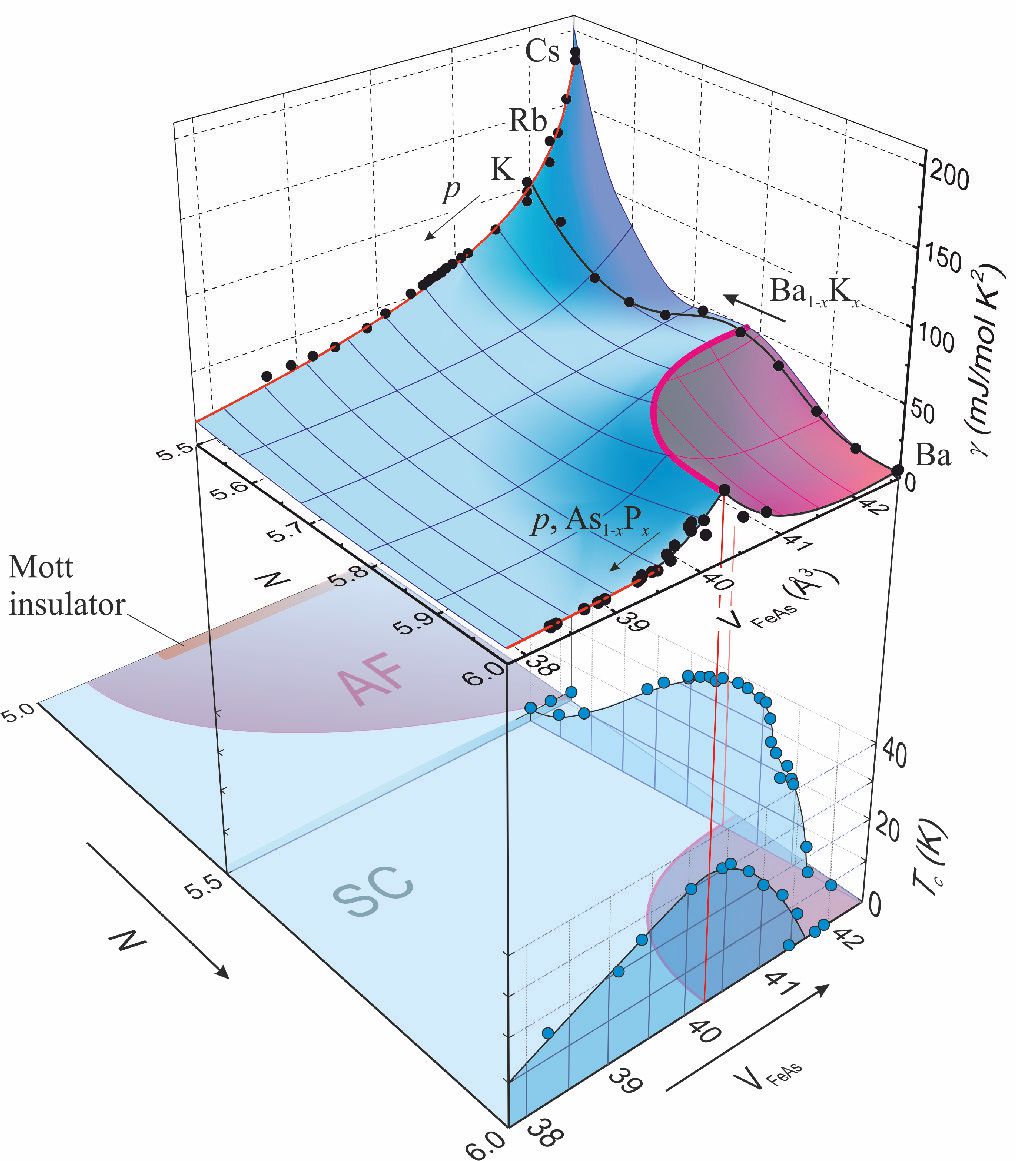}%
\caption{\label{fig:base} (Color online) 
Evolution of $\gamma$ as a function of $V_{\text{FeAs}}$ and hole doping expressed as filling $N$ of the Fe 3$d$ states. Apart from high-pressure data of KFe$_2$As$_2$ \cite{Taufour2014,Ying2015}, measurements of Ba$_{1-x}$K$_x$Fe$_2$As$_2$ \cite{Storey2013,Boehmer2015} and BaFe$_2$(As$_{1-x}$P$_x$)$_2$ \cite{Shibauchi2014,Walmsley2013,Kasahara2010,Klintberg2010} have been included. The $\gamma$ values of BaFe$_2$(As$_{1-x}$P$_x$)$_2$ were estimated following Ref.\,\onlinecite{Walmsley2013}. The base plane shows a tentative phase diagram of a five-orbital Hubbard model with combined $(U,J)$ interactions leading to a Mott-insulating state at $N=5$ \cite{Yu2013II}.
}
\end{figure}

The specific role of the Fe $3d$ levels for the electronic correlations in Fe-based  superconductors reflects the Coulomb (Hubbard and Hund) interactions as well as the small crystal-electric-field splittings \cite{deMedici2014,Georges2013,Yu2013II}. To better understand the observed mass enhancement, we study the electron correlation effects in a multiorbital Hubbard model for $A$122 using a $U(1)$ slave-spin mean-field theory \cite{Yu2012}. The details of the model and the method can be found in the Supplementary Material. The $A$122 system corresponds to a $3d$-electron filling of $N=5.5$ per Fe atom. For this filling, we identify a strongly correlated regime for a range of realistic values for $U$ and $J$ (Fig.~S5 of Ref.\,\onlinecite{Suppl}). In this regime, the quasiparticle spectral weights in all five Fe $3d$ orbitals are substantially reduced, and a strong orbital dependence arises. As shown in Fig.~S6, the quasiparticle spectral weight of the $d_{xy}$ orbital is most strongly reduced; correspondingly, its mass enhancement is the largest. We further calculate the Sommerfeld coefficient $\gamma$. As shown in Fig.~\ref{fig:all}(b), the calculated $\gamma$ values for the three compounds have a magnitude similar to the experimental values. However, for fixed values of the interactions, $\gamma$ does not show a strong increase across the K through Rb to Cs series. We attribute this missing contribution to an additional component of $\gamma$ arising from an AF quantum criticality.

The emerging picture is summarized in Fig.\,\ref{fig:base} which depicts $\gamma$ in the $(N,V_{\text{FeAs}})$ plane. The $\gamma$ values of the series Ba$_x$K$_{1-x}$Fe$_2$As$_2$ \cite{Storey2013}, where $N=6-x/2$ varies from $N=5.5$ to 6, contrast with those of BaFe$_2$(As$_{1-x}$P$_x$)$_2$, where $N=6$ and $V_{\text{FeAs}}$ changes significantly \cite{Shibauchi2014,Walmsley2013}. 
Both series exhibit a maximum $\gamma_{max}$ close to the onset of AF order suggesting an underlying QCP. However, the highest $\gamma$ value is found for $A$122 where $T_c$ vanishes. 
We note that the landscape of $\gamma(N,V_{\text{FeAs}})$ is compatible with a line of $\gamma_{max}$ connecting the values of both sides. Figure~\ref{fig:base} strongly suggests that the QCP identified by our measurements would be most naturally associated with an AF order related to the $N=5$ Mott-insulating phase \cite{Yu2013II}.

The combined effect of Hund's rule coupling and Coulomb exchange interaction is considerably intensified by reducing the band width. This is exemplified by BaFe$_2$(As$_{1-x}$P$_x$)$_2$ and $A$122. While in BaFe$_2$(As$_{1-x}$P$_x$)$_2$ the hybridization changes due to longer Fe-As bonds, in the $A$122 series only the Fe-Fe distances are widened \cite{Suppl,Kasahara2010}. Remarkably, the largest $\gamma$ values are found for $A$122 with a divergent trend towards $A$~=~Cs. These high values do not find any correspondence in the superconducting properties. The $T_c$ values are one order of magnitude smaller than those of Ba$_x$K$_{1-x}$Fe$_2$As$_2$ and BaFe$_2$(As$_{1-x}$P$_x$)$_2$, and even decrease with increasing quasiparticle masses. In fact, it seems that for $A$122 strong correlations and SC compete with each other which might be the case on rather general grounds \cite{Monthoux1994,Moriya2000}. 
We speculate that this series would be a candidate for the special situation where a QCP exactly coincides with the onset of SC. This could provide a new insight into the physics of iron pnictides close to the $N=5$ limit which should be tested by investigating Cs122 films under tensile biaxial strain.
 
\begin{acknowledgments}
We thank Y. Mizukami, H. Ikeda, T. Terashima, F. Hardy, A. E. B\"ohmer, C. Meingast, J. Schmalian, L. de' Medici, M. Capone, M. Grosche, and J. Wosnitza for valuable discussions. This work has been supported by Deutsche Forschungsgemeinschaft in the frame of FOR960 (Quantum Phase Transitions) and the Japan-Germany Research Cooperative Program, KAKENHI from JSPS and Project No. 56393598 from DAAD. The work was in part supported by U.S. DOE at LANL under Contract No. DE-AC52-06NA25396 and the DOE Office of Basic Energy of Sciences (J.-X. Z.), by  the  National  Science Foundation of China Grant number 11374361, the Fundamental Research Funds for the Central Universities and the Research Funds of Renmin University of China (R.Y.), and by the NSF and the Robert A.\ Welch Foundation Grant No.\ C-1411 (Q.S.). Q.S. also acknowledges the support of the Alexander von Humboldt Foundation through a Humboldt Research Award and the hospitality of the Karlsruhe Institute of Technology. H.v.L. and Q.S. enjoyed the hospitality of the Aspen Center for Physics when finalizing the manuscript (supported by NSF grant PHY-1066293). 
\end{acknowledgments}

%

\end{document}


\title{SUPPLEMENTARY MATERIAL -- \\
Quantum criticality in $A$Fe$_2$As$_2$ with $A$~= K, Rb, and Cs suppresses superconductivity}

\author{Felix Eilers}
\author{Kai Grube}
\author{Diego A. Zocco}
\author{Thomas Wolf}
\author{Michael Merz}
\author{Peter Schweiss}
\author{Rolf Heid}
\author{Robert Eder}
\affiliation{Institut f\"ur Festk\"orperphysik, Karlsruher Institut f\"ur Technologie, 76021 Karlsruhe, Germany}
\author{Rong Yu}
\affiliation{Department of Physics, Renmin University of China, Beijing 100872, China}
\author{Jian-Xin Zhu}
\affiliation{Theoretical Division, Los Alamos National Laboratory, Los Alamos, New Mexico 87545, USA}
\author{Qimiao Si}
\affiliation{Department of Physics and Astronomy, Rice University, Houston, TX 77005, USA}
\author{Takasada Shibauchi}
\affiliation{Department of Advanced Materials Science, University of Tokyo, Kashiwa, Chiba 277-8561, Japan}
\affiliation{Department of Physics, Kyoto University, Sakyo-ku, Kyoto 606-8502, Japan}
\author{Hilbert v. L\"ohneysen}
\affiliation{Institut f\"ur Festk\"orperphysik, Karlsruher Institut f\"ur Technologie, 76021 Karlsruhe, Germany}
\affiliation{Physikalisches Institut, Karlsruhe Institut f\"ur Technologie, 76049 Karlsruhe, Germany}

\maketitle

\setcounter{figure}{0}
\makeatletter
\renewcommand{\thefigure}{S\@arabic\c@figure}

\section{Experimental details}
Single crystals of RbFe$_2$As$_2$ and CsFe$_2$As$_2$ were grown from arsenic-rich flux (composition $0.40:0.05:0.55$ $A$:Fe:As) in alumina crucibles, as reported in Ref.\,\cite{Zocco2013} for KFe$_2$As$_2$. The crucibles were sealed in an iron tube filled with argon gas. Maximum temperatures were 980 and $950\,^\circ$C, respectively for $A$~= Rb, Cs, minimum temperatures 690 and $800\,^\circ$C, with cooling rates of $0.76$ and $0.2$ to $0.25\,^\circ$C/hour. The crystals were annealed in situ, directly after the growth, for one day each at 450, 400, and $350\,^\circ$C. The crystal structure was examined with four-circle x-ray diffraction at room temperature. Structural refinement confirmed the space group to be $I4/mmm$ and the composition to be stoichiometric within the error of the experiment (1-2\,\%). The lattice parameters were determined to be $a=3.872(4)$\,{\AA}, $c=14.46(5)$\,{\AA}, $z=0.3475(3)$ for RbFe$_2$As$_2$ and $a=3.902(2)$\,{\AA}, $c=15.14(2)$\,{\AA}, $z=0.3416(4)$ for CsFe$_2$As$_2$. 

\begin{figure}[t!]
\includegraphics[width=8.6cm]{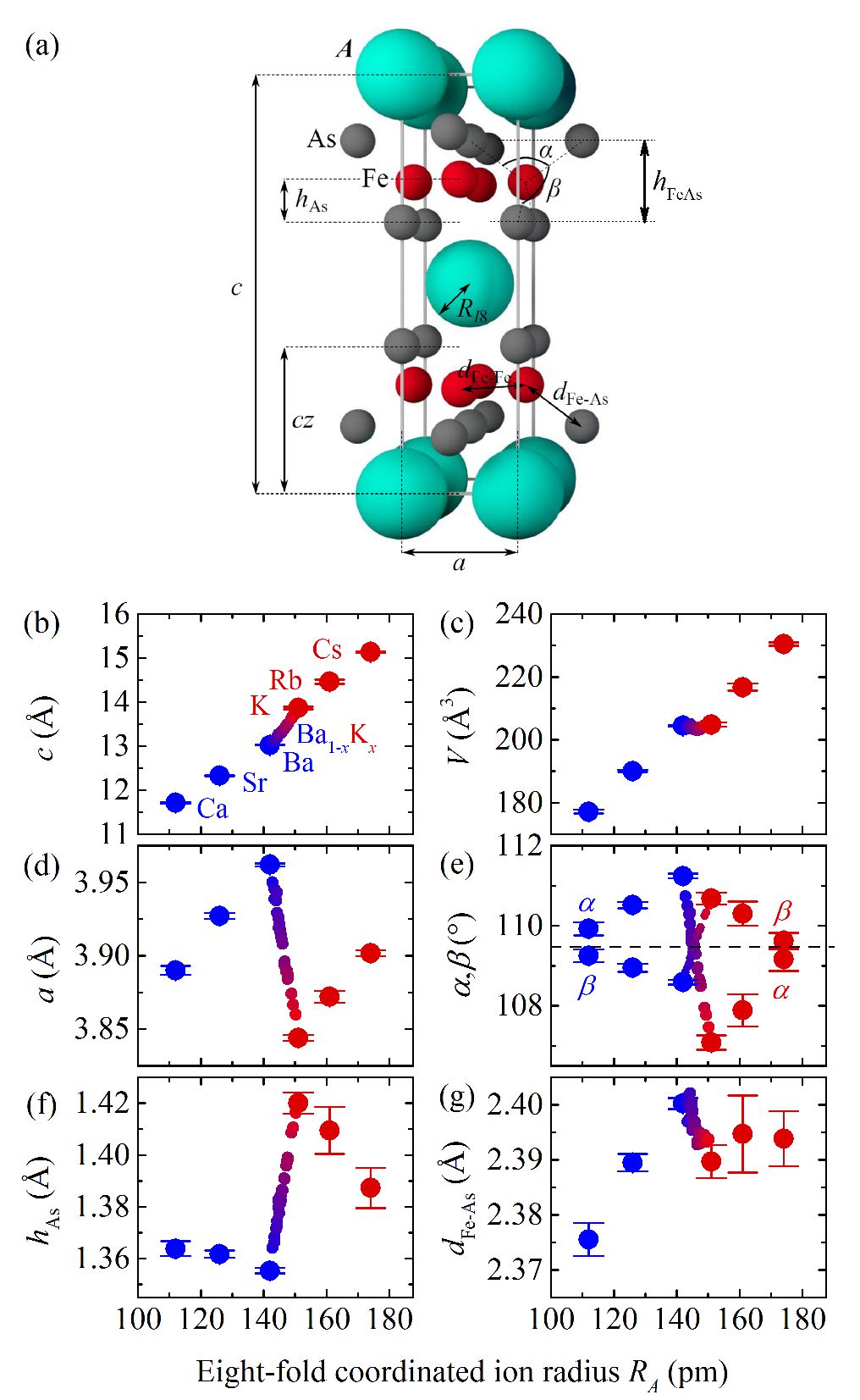}%
\caption{\label{fig:struct} (Color online) (a) Conventional tetragonal unit cell of $A$Fe$_2$As$_2$. (b-g) Structural parameters for $A$~= Ca, Sr, Ba, K, Rb, and Cs plotted against the eight-fold coordinated ion radius $R_A$.
}
\end{figure}

Figure \ref{fig:struct} shows the structural parameters of RbFe$_2$As$_2$ and CsFe$_2$As$_2$ in comparison with CaFe$_2$As$_2$, SrFe$_2$As$_2$, BaFe$_2$As$_2$, and KFe$_2$As$_2$. The $c$ parameter increases steadily with the eight-fold coordinated ion-radius of the alkaline or alkaline earth atom $R_A$. The tetrahedron of arsenic atoms surrounding the iron atoms is compressed along the $c$-direction for $AE=$ Ca, Sr, Ba and elongated along $c$ for $A=$ K, Rb, Cs. The crystal structure of CsFe$_2$As$_2$ is very similar to that of Ba$_{0.5}$K$_{0.5}$Fe$_2$As$_2$. The iron-iron distance $d_{\mathrm{Fe-Fe}} = a/\sqrt{2}$ as well as the iron-arsenic distance $d_{\mathrm{Fe-As}}$ change only little in the $A$~= K, Rb, Cs series. 

The magnetostriction and thermal expansion have been measured using a parallel-plate capacitance dilatometer operated in a dilution refrigerator, at temperatures between 20\,mK and 10\,K and at magnetic fields up to 14\,T. Changes in length of the crystals were measured both along the $c$ and the $a$ direction for each direction of the applied magnetic field, along $c$ and along $a$. Field sweeps were recorded using a rate of $0.02$\,T/min and $0.1$\,T/min for $B \parallel c$ and $a$, respectively. The thermal expansion and magnetostriction are measured with a negligible stress ($<1\,$bar) exerted to the sample by the spring suspension of the dilatometer. Therefore, the obtained stress and strain dependences only reflect changes produced by infinitesimal uniaxial pressures. Thus, for a tetragonal system in the basal plane, $\alpha_a$ is equal to $\alpha_b$ and $\partial \gamma/\partial \epsilon_a=\partial \gamma/\partial \epsilon_b$ or $\partial T_c/\partial \epsilon_a=\partial T_c/\partial \epsilon_b$. 

\begin{figure}[t!]
\includegraphics{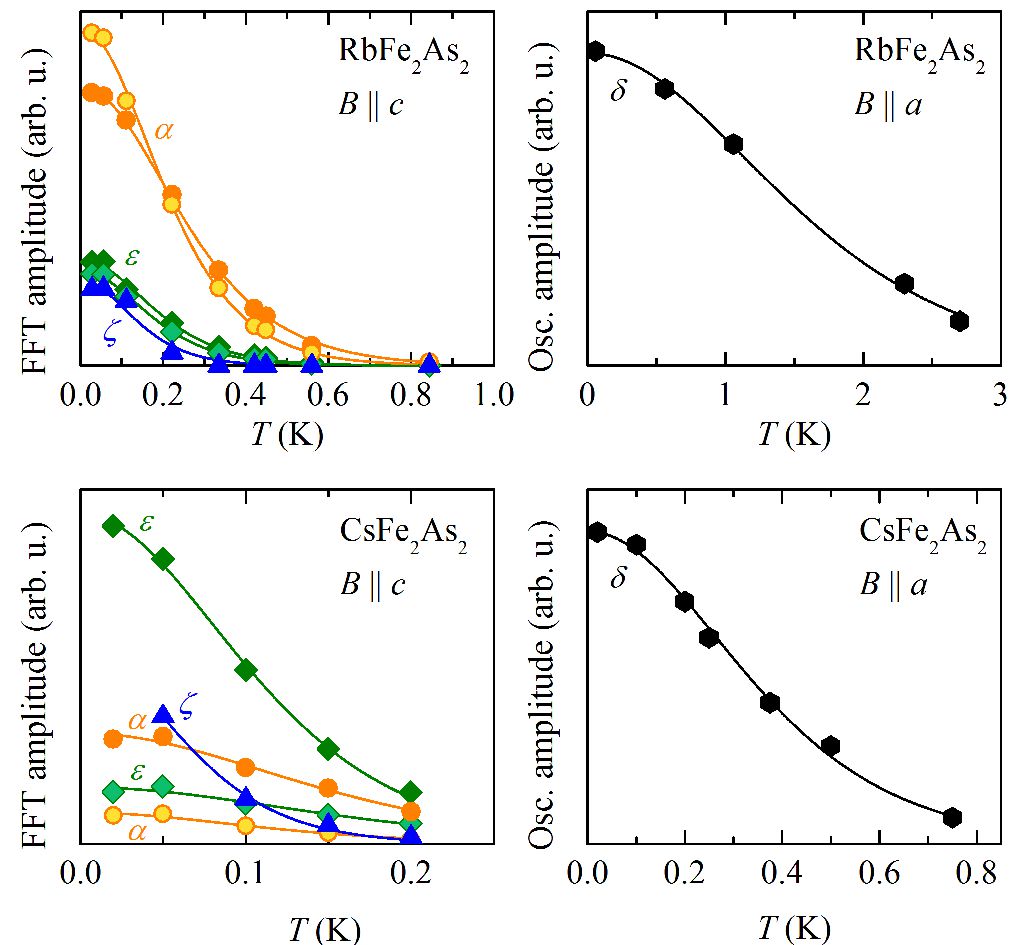}%
\caption{\label{fig:fits}Fourier amplitudes ($B \parallel c$) and oscillation amplitudes ($B \parallel a$) plotted against temperature. Dashed lines are fits of the temperature factor.}
\end{figure}

To analyze the quantum oscillation observed in the magnetostriction, the change in capacitance was converted to a change in length, the derivative with respect to the magnetic field was calculated and the oscillatory part was extracted by subtracting a polynomial of low order. The fundamental frequencies were extracted from the Fourier spectra of the oscillatory part of the magnetostriction coefficients $\lambda_i \equiv L_i^{-1} \partial L_i / \partial B$, where $L_i$ is the length of the crystal in the $i=a,c$ direction and $B=\mu_0 H$ is the applied magnetic field. Values of the effective quasiparticle mass $m^*$ have been inferred from the temperature dependence of the quantum oscillations by using the Lifshitz-Kosevich formula \cite{Shoenberg1984,Zocco2014}. Figure~\ref{fig:fits} displays the fits for the Fermi-surface sheets obtained for $B \parallel a$ and $c$. 

\section{Theoretical caculations}
\subsection{Elastic constants of $A$Fe$_2$As$_2$ ($A$~= K, Rb, Cs)}
To obtain estimates of the elastic constants we performed ab-initio calculations of electronic and vibrational properties of $A$Fe$_2$As$_2$ ($A$~= K, Rb, CS) in the framework of the generalized gradient approximation using a mixed-basis pseudopotential method \cite{meyer}. Phonon dispersions and corresponding interatomic force constants were calculated via density functional perturbation theory \cite{heid99}. The elastic constants were then derived via the method of long waves, which relates the elastic constants to sums over force constants weighted by the bond vector \cite{born54}. All calculations used experimental lattice structures, and employed normconserving pseudopotentials with a plane-wave cutoff of 22\,Ry, augmented by local functions at the sites of both Fe and the alkaline metals.  Brillouin zone summations were done with a Gaussian broadening technique using a broadening of 0.1\,eV and 40 wave vector points in the irreducible part of the Brillouin zone.  Additional technical details can be found in Ref.~\cite{mittal09,reznik09}, where the same approach has been applied to studies of the vibrational properties of CaFe$_2$As$_2$ and BaFe$_2$As$_2$. We obtain the following sets of elastic constants $c_{ij}$ with $i,j=1\cdots 3$ and bulk moduli $B_T$ for KFe$_2$As$_2$ (in GPa):
\begin{equation}
	c_{ij} = 
	\begin{pmatrix}
		79.7 & 46.6 & 38.4 \\
		     & 79.7 & 38.4 \\
		     &      & 45.1 \\ 
	\end{pmatrix}, B_T=43.7\,\text{GPa},
\label{elast_K122}
\end{equation}
for RbFe$_2$As$_2$ (in GPa):
\begin{equation}
	c_{ij} = 
	\begin{pmatrix}
		78.3 & 40.2 & 28.4 \\
		     & 78.3 & 28.4 \\
		     &      & 53.8 \\ 
	\end{pmatrix}, B_T=42.3\,\text{GPa},
\label{elast_Rb122}
\end{equation}
and for CsFe$_2$As$_2$ (in GPa):
\begin{equation}
	c_{ij} = 
	\begin{pmatrix}
		84.4 & 45.9 & 39.5 \\
		     & 84.4 & 39.5 \\
		     &      & 61.2 \\ 
	\end{pmatrix}, B_T=51.3\,\text{GPa}.
\label{elast_Cs122}
\end{equation}
The measured bulk modulus of KFe$_2$As$_2$ amounts to $40\pm 1$\,GPa \cite{Tafti2014}.

\subsection{Electron correlations in $A$Fe$_2$As$_2$ ($A$~= K, Rb, Cs)}
To consider the correlation effects in theory, we study a multiorbital Hubbard model
for the $A$Fe$_2$As$_2$ ($A$~= K, Rb, Cs) system. The Hamiltonian reads
\begin{equation}
 \label{Eq:Ham_tot} H=H_0 + H_{\mathrm{int}}.
\end{equation}
$H_0$ contains the tight-binding parameters among the multiple orbitals of $A$Fe$_2$As$_2$.
\begin{equation}
 \label{Eq:Ham_0} H_0=\frac{1}{2}\sum_{ij\alpha\beta\sigma} t^{\alpha\beta}_{ij} d^\dagger_{i\alpha\sigma} d_{j\beta\sigma} + \sum_{i\alpha\sigma} (\Delta_\alpha-\mu) d^\dagger_{i\alpha\sigma} d_{i\alpha\sigma},
\end{equation}
where $d^\dagger_{i\alpha\sigma}$ creates an electron in orbital $\alpha$ with spin $\sigma$ at site $i$, $\Delta_\alpha$
is the on-site energy reflecting the crystal field splitting, and $\mu$ is the chemical potential. 
The details of the tight-binding parameters will be discussed in the next section of this Supplementary Material.
$H_{\rm{int}}$ contains on-site Hubbard interactions
\begin{eqnarray}
 \label{Eq:Ham_int} H_{\rm{int}} &=& \frac{U}{2} \sum_{i,\alpha,\sigma}n_{i\alpha\sigma}n_{i\alpha\bar{\sigma}}\nonumber\\
 &&+\sum_{i,\alpha<\beta,\sigma} \left\{ U^\prime n_{i\alpha\sigma} n_{i\beta\bar{\sigma}}\right. 
 + (U^\prime-J) n_{i\alpha\sigma} n_{i\beta\sigma}\nonumber\\
&&\left.-J(d^\dagger_{i\alpha\sigma}d_{i\alpha\bar{\sigma}} d^\dagger_{i\beta\bar{\sigma}}d_{i\beta\sigma}
 -d^\dagger_{i\alpha\sigma}d^\dagger_{i\alpha\bar{\sigma}}
 d_{i\beta\sigma}d_{i\beta\bar{\sigma}}) \right\}.
\end{eqnarray}
where $n_{i\alpha\sigma}=d^\dagger_{i\alpha\sigma} d_{i\alpha\sigma}$. In this model,
$U$, $U^\prime$, and $J$ respectively denote the intraorbital repulsion, the interorbital repulsion,
and the Hund's rule exchange coupling. We take $U^\prime=U-2J$ as is standard \cite{Castellani78}.

\begin{figure}[t!]
\includegraphics[width=8.6cm]{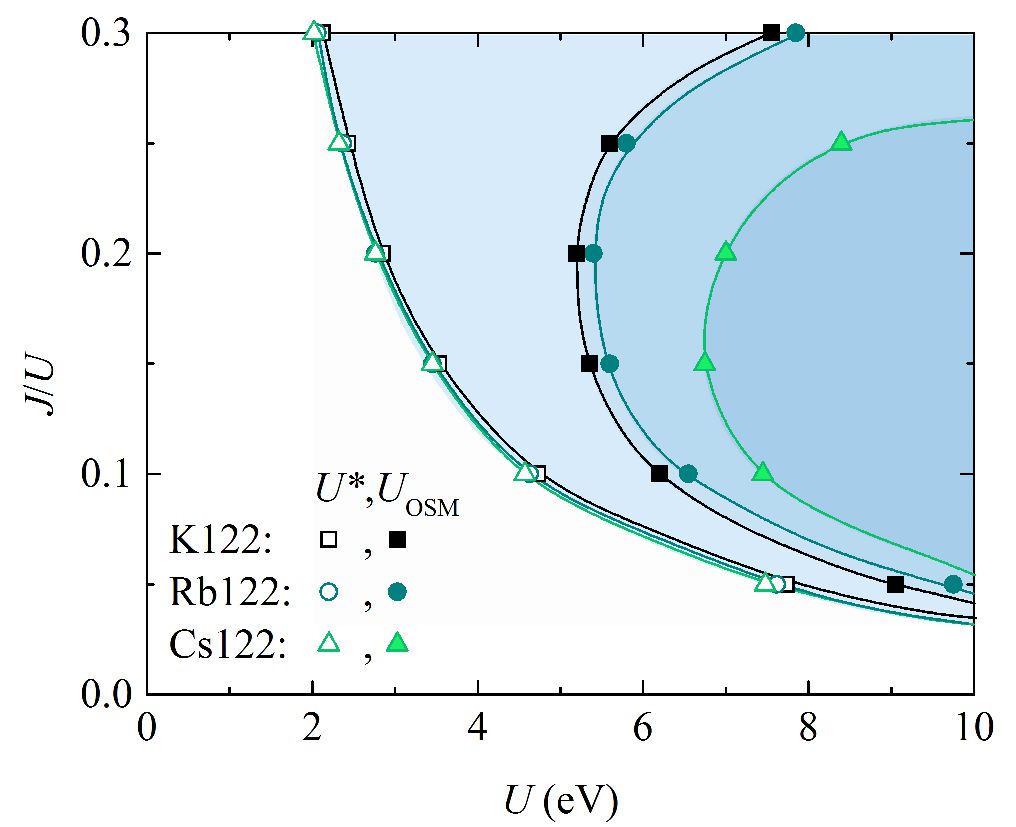}%
\caption{\label{fig:S1} (Color online) Ground-state phase diagram of KFe$_2$As$_2$, RbFe$_2$As$_2$, and CsFe$_2$As$_2$ in the $J/U$-$U$ plane from slave-spin mean-field theory. The open and closed symbols show the crossover at $U^\star$, and the critical $U$ value for the orbital-selective Mott transition, $U_{OSM}$, respectively.
}
\end{figure}

\begin{figure}[t]
\includegraphics[width=8.6cm]{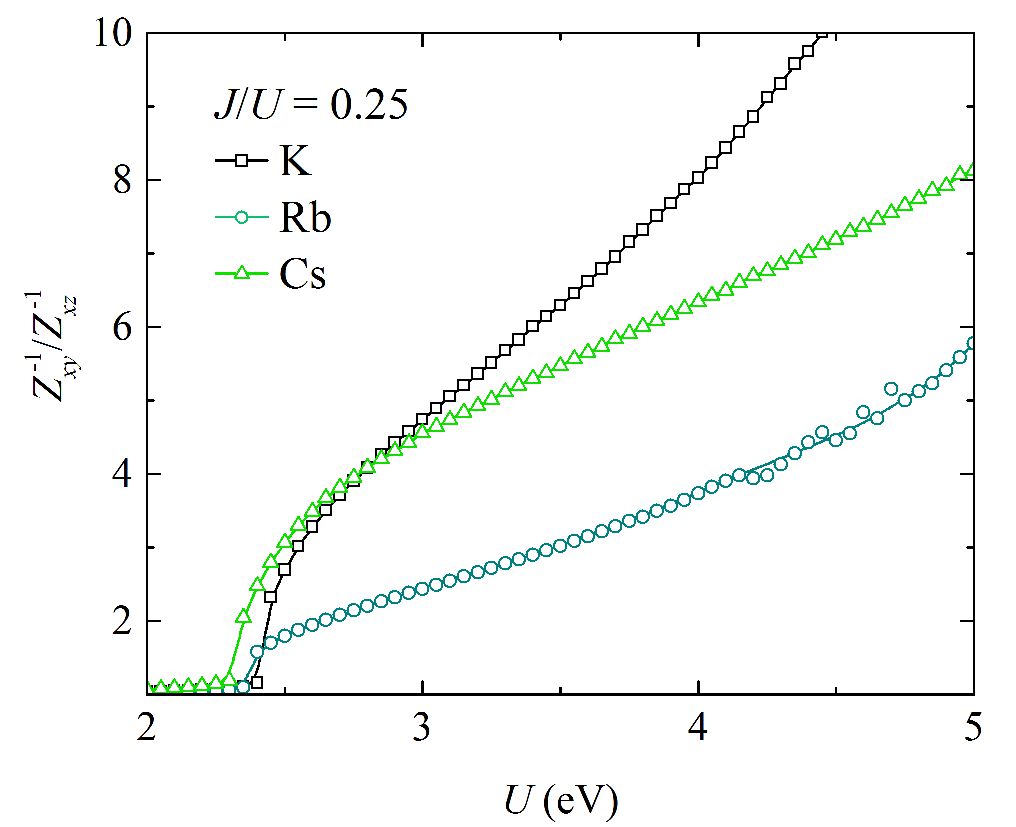}%
\caption{\label{fig:S2} (Color online) Evolution of the ratio $Z^{-1}_{xy}/Z^{-1}_{xz}$ with $U$ at $J/U=0.25$ in the models for KFe$_2$As$_2$, RbFe$_2$As$_2$, and CsFe$_2$As$_2$. $Z^{-1}_{xy}/Z^{-1}_{xz}$ is proportional to the ratio of mass enhancement of $d_{xy}$ orbital to $d_{xz}$ orbital. For $U\gtrsim 2.5$ eV, a wide regime with strong orbital selectivity exists for the three compounds, with the heaviest orbital to be the 3$d_{xy}$.
}
\end{figure}

\begin{figure}[t]
\includegraphics[width=8.6cm]{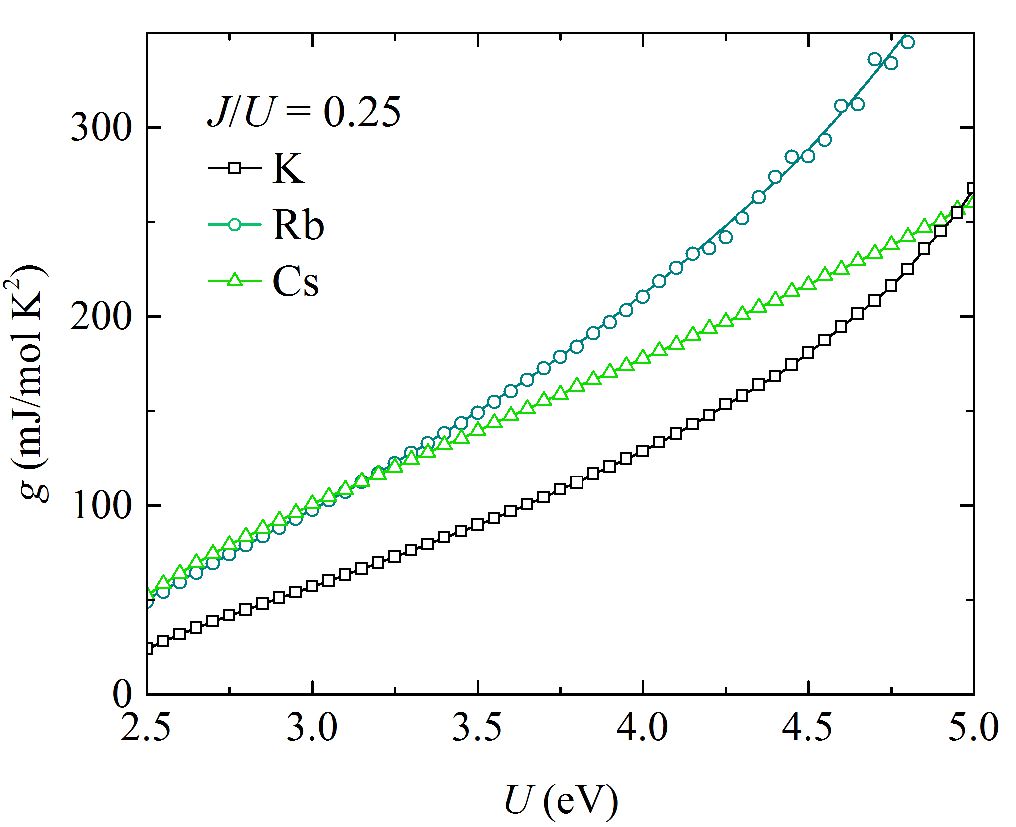}%
\caption{\label{fig:S3} (Color online) The evolution of $g$ with $U$. The definition of $g$ is given in Eq.~\ref{Eq:g}. $g$ behaves in a similar way to $\gamma$ in the regime with strong electron correlations.
}
\end{figure}

\begin{figure}[b]
\includegraphics[width=8.6cm]{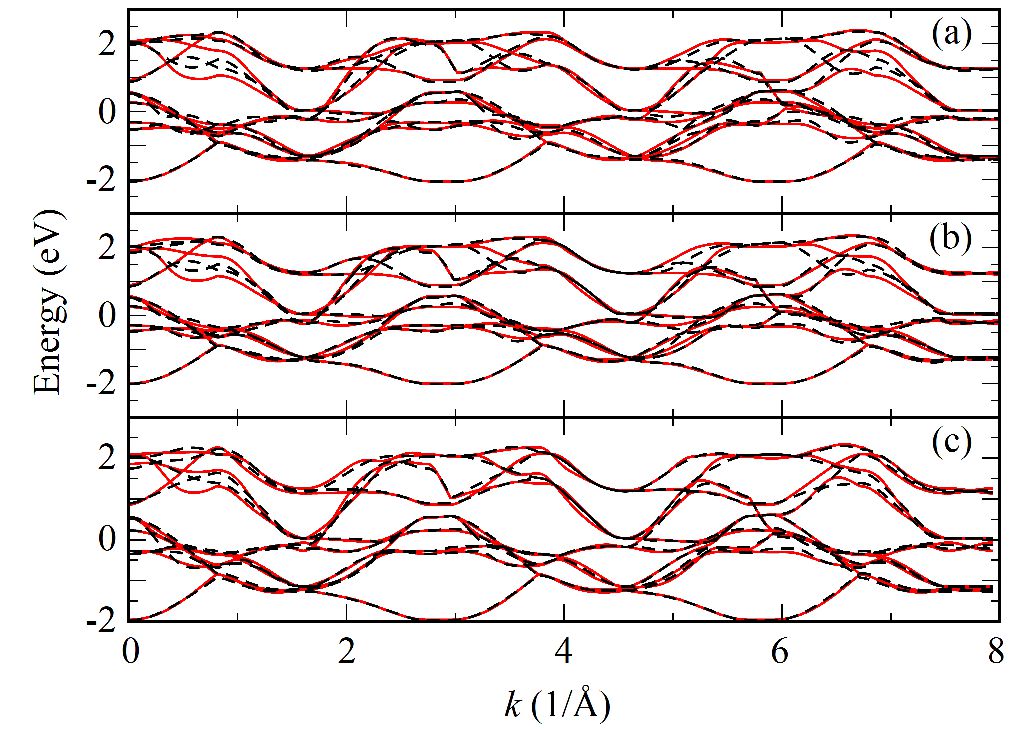}%
\caption{\label{fig:S4} (Color online) The five-orbital tight-binding fit (dashed black lines) of the 10 Wannier fit (solid red lines) for KFe$_2$As$_2$ (a), RbFe$_2$As$_2$ (b), CsFe$_2$As$_2$ (c). All energies are measured with respect to the Fermi energy. The wave vector $k$-path is chosen to be along the high symmetry points.
}
\end{figure}

We study the model via a U(1) slave-spin approach~\cite{Yu2012}. Within this theory, the effect of electron correlations can be characterized by the orbital-resolved quasiparticle spectral weight, $Z_\alpha$, which we will calculate in the paramagnetic phase at the electron filling of $n=5.5$ electrons per Fe ion. The results are summarized in the ground-state phase diagram given in Fig.~\ref{fig:S1}. Because $N$ is not an integer, no metal-to-insulator transition is observed; the system is always metallic for any $U$ and $J$ values. We do however find, for a fixed $J$ and with increasing $U$, a crossover at $U^\star$ from a weakly correlated regime (with $Z_\alpha\sim1$) to a regime with strong correlation ($Z_\alpha\ll1$). Further increasing $U$ (or $J$ for a fixed $U$) drives the system into an orbital-selective Mott phase (OSMP) in which the 3$d_{xy}$ orbital of Fe is Mott localized (with $Z_{xy}=0$) while the other 3$d$ orbitals are still itinerant (with $Z>0$). We find that the critical value of the orbital-selective Mott transition (OSMT), $U_{OSM}$, shows strong material dependence from K to Cs, and in general $U_{OSM}\gtrsim5.5$\,eV, while $U^\star$ is comparable for all the three compounds, and can be as low as $\sim 2.3$\,eV for $J/U=0.25$. Therefore, there is a large regime in the phase diagram that shows strong orbital selectivity, as demonstrated in Fig.~\ref{fig:S2}. It is also an interesting observation that similar strong orbital selectivity shown in the phase diagram for these heavily hole doped iron pnictides also exists in the phase diagrams for some parent and electron doped iron pnictides and iron selenides~\cite{Yu2012,Yu2013I}.

The effect of strong electron correlations is also manifested through the Sommerfeld coefficient $\gamma$. We have calculated $\gamma$ by taking the second derivative of the free energy with respect to temperature, namely, $\gamma=-\partial^2f/\partial T^2$. The results for the three compounds are shown in Fig.\,3(b) of the main text. We find that for each of the three compounds, $\gamma$ is enhanced in the regime of phase diagram that exhibits strong electron correlations. Remarkably, within a certain range of $U$ and $J$ ($2.5$ eV $\lesssim U \lesssim 5$ eV for $J/U=0.25$, for example), the calculated values of $\gamma$ are about $100\sim200$ mJ/mol K$^2$, which are of the same order as the experimental values for these compounds. 

In the slave-spin approach, an analytical expression of $\gamma$ is difficult to obtain due to the multiorbital nature. However, we can gain insights into the behavior of $\gamma$ from an approximate quantity, $g$. 
In the regime of phase diagram showing strong orbital selectivity, the inter-orbital correlations 
are suppressed by the combined effects of Hund's coupling and crystal level splitting. Therefore, to a good 
approximation, each orbital contributes to $\gamma$ independently. We therefore define 
\begin{equation}\label{Eq:g}
g=\sum_\alpha \mathfrak{N}_\alpha(E_F)/Z_\alpha,
\end{equation}
where $\mathfrak{N}_\alpha(E_F)$ is the electron density of states at the Fermi energy projected onto orbital $\alpha$ in the noninteracting limit. Fig.~\ref{fig:S3} shows the evolution of $g$ with $U$, calculated with the same model parameters as in Fig.~3(b) of the main text. The qualitatively similar behavior of $g$ and $\gamma$ implies the same underlying physics. From the definition of $g$, we see that the enhancement of $g$ (and $\gamma$) originates from the mass enhancement of each orbital, which is proportional to $1/Z_\alpha$. Due to the strong orbital selectivity (Fig.~\ref{fig:S2}), the 3$d_{xy}$ orbital is the heaviest, and contributes to $g$ (and $\gamma$) the most. At the OSMT, the $d_{xy}$ orbital is fully decoupled from other orbitals, and $Z_{xy}=0$. As a consequence, $g$ and $\gamma$ diverge as the OSMT is approached. This provides one way to understand the strongly enhanced $\gamma$.

However, for a fixed set of $U$ and $J$ values, $\gamma$ does not increase monotonically from K to Cs. As discussed in the main text, this implicates an additional contribution to $\gamma$ associated with AF quantum criticality. A natural candidate is a quantum critical point (QCP) for an AF ordering associated with the Mott insulating state at $N=5$; the latter has been advanced in Ref.~\onlinecite{Yu2013II} and illustrated in Fig.\,4 of the main text.

\subsection{DFT calculations and tight-binding parameters for $A$Fe$_2$As$_2$ ($A$~= K, Rb, Cs)} 
We have performed band-structure calculations for KFe$_2$As$_2$, RbFe$_2$As$_2$ and CsFe$_2$As$_2$ based on the generalized gradient approximation. The full-potential linearized augmented plane wave (FP-LAPW) method as implemented in the WIEN2K code~\cite{PBlaha:2001} is used. We then follow the procedure suggested by Graser {\em et al.}~\cite{SGraser:2010} to fit the Wannierized bands~\cite{NMarzari:1997,ISouza:2001} with a five-orbital tight-binding Hamiltonian~\cite{SGraser:2009}, by unfolding the Brillouin zone with two Fe sites per unit cell to the Brillouin zone corresponding to one Fe site per unit cell. In this procedure, an interface~\cite{JKunes:2010} between the WIEN2k code and the wannier90 code~\cite{AMostofi:2008} has also been employed to disentangle the bands. We show in Fig.~\ref{fig:S4} the comparison of the tight-binding fit to those obtained from the GGA-based Wannier orbitals for the three 122 compounds. The corresponding tight-binding model parameters are listed in Tables S1-S3.

\begin{widetext}

\begin{minipage}{\linewidth}
\begin{center}
\begin{tabular}{cccccccccccc}
  \hline
  \hline
    & $\alpha=1$ & $\alpha=2$ & $\alpha=3$ & $\alpha=4$ & $\alpha=5$ &   & & & & &  \\ \hline
  $\epsilon_\alpha$ & 0.27530 & 0.27530 & -0.37164 & 0.69843 & 0.1381 &  & \\ \hline\hline
  $t^{\alpha\alpha}_\mu$ & $\mu=x$ & $\mu=y$ & $\mu=xy$ & $\mu=xx$ & $\mu=xxy$ & $\mu=xyy$ & $\mu=xxyy$ & $\mu=z$ & $\mu=xz$ & $\mu=xxz$ & $\mu=xyz$ \\ \hline
  $\alpha=1$ & -0.08418 & -0.19954 & 0.06963 & 0.24664 & -0.07912 & 0.03223 & 0.01942 &  & -0.00580 & 0.04127 & \\ \hline
  $\alpha=3$ & 0.36702 &  & 0.02663 & -0.08301 &  &  & & & & &  \\ \hline
  $\alpha=4$ & 0.15386 &  & -0.05682 & 0.10419 & -0.01544 &  & -0.06894 & 0.09484 & 0.06175 & & 0.02122 \\ \hline
  $\alpha=5$ & -0.02496 &  & -0.12021 &  & -0.01013 &  & 0.02655 & 0.03076 & -0.02373 & & \\ \hline\hline
  $t^{\alpha\beta}_\mu$ & $\mu=x$ & $\mu=xy$ & $\mu=xxy$ & $\mu=xxyy$ & $\mu=z$ & $\mu=xz$ & $\mu=xyz$ & $\mu=xxyz$ & & & \\ \hline
  $\alpha\beta=12$ &  & -0.14893 & -0.06209 & -0.03947 &  &  & 0.01886 & & & & \\ \hline
  $\alpha\beta=13$ & -0.43065 & -0.00736 & -0.05048 &  &  &  & & & & & \\ \hline
  $\alpha\beta=14$ & 0.21335 & -0.06953 & 0.05399 &  &  & -0.00344 & 0.00223 & -0.00263 & & & \\ \hline
  $\alpha\beta=15$ & -0.19991 & -0.14116 &  & 0.00000 &  &  & -0.00153 & & & & \\ \hline
  $\alpha\beta=24$ &  &  &  &  &  & 0.03195 &  & 0.01817 & & & \\ \hline
  $\alpha\beta=34$ &  &  & -0.05942 &  &  &  & & & & & \\ \hline
  $\alpha\beta=35$ & -0.22330 &  & 0.05165 &  &  &  & & & & & \\ \hline
  $\alpha\beta=45$ &  & -0.30785 &  & 0.00160 & -0.08049 & 0.01885 & & & & & \\ \hline
  \hline
\end{tabular}
\par
\end{center}
\noindent
{\bf Table S}1.
 Tight-binding parameters of the five-orbital model for \text{KFe$_{2}$As$_2$}. Here we use the same notation as in Ref.~\onlinecite{SGraser:2010}.
The orbital indeces $\alpha=$1,2,3,4,5 correspond to $d_{xz}$, $d_{yz}$,
$d_{x^2-y^2}$, $d_{xy}$, and $d_{3z^2-r^2}$ orbitals, respectively.
The units of the parameters are eV.
\end{minipage}


\begin{minipage}{\linewidth}
\begin{center}
\begin{tabular}{cccccccccccc}
  \hline
  \hline
    & $\alpha=1$ & $\alpha=2$ & $\alpha=3$ & $\alpha=4$ & $\alpha=5$ &   & & & & &  \\ \hline
  $\epsilon_\alpha$ & 0.32345 & 0.32345 & -0.40194  & 0.71731 & 0.14450 &  & \\ \hline\hline
  $t^{\alpha\alpha}_\mu$ & $\mu=x$ & $\mu=y$ & $\mu=xy$ & $\mu=xx$ & $\mu=xxy$ & $\mu=xyy$ & $\mu=xxyy$ & $\mu=z$ & $\mu=xz$ & $\mu=xxz$ & $\mu=xyz$ \\ \hline
  $\alpha=1$ & -0.06985 & -0.22598 & 0.08377 & 0.24043 & -0.06168 & 0.02105 & 0.01175 &  & -0.00650 & 0.03774 & \\ \hline
  $\alpha=3$ & 0.35885 &  & 0.05108 & -0.09566 &  &  & & & & &  \\ \hline
  $\alpha=4$ & 0.13022 &  & -0.07484 & 0.10261 & -0.00952 &  & -0.05511 & 0.09126 & 0.05656 & & 0.01980 \\ \hline
  $\alpha=5$ & -0.02614 &  & -0.12488 &  & -0.00458 &  & 0.03092 & 0.02191 & -0.01853 & & \\ \hline\hline
  $t^{\alpha\beta}_\mu$ & $\mu=x$ & $\mu=xy$ & $\mu=xxy$ & $\mu=xxyy$ & $\mu=z$ & $\mu=xz$ & $\mu=xyz$ & $\mu=xxyz$ & & & \\ \hline
  $\alpha\beta=12$ &  & -0.18759 & -0.05309 & -0.03053 &  &  & 0.01131 & & & & \\ \hline
  $\alpha\beta=13$ & -0.29129 & -0.03934 & -0.00812 &  &  &  & & & & & \\ \hline
  $\alpha\beta=14$ & 0.16066 & -0.10845 & 0.06566 &  &  & 0.00170 & 0.00105 & -0.00054 & & & \\ \hline
  $\alpha\beta=15$ & -0.16746 & -0.15103 &  & -0.00000 &  &  & -0.01179 & & & & \\ \hline
  $\alpha\beta=24$ &  &  &  &  &  & 0.02094 &  & 0.02331 & & & \\ \hline
  $\alpha\beta=34$ &  &  & -0.03127 &  &  &  & & & & & \\ \hline
  $\alpha\beta=35$ & -0.35706 &  & -0.02342 &  &  &  & & & & & \\ \hline
  $\alpha\beta=45$ &  & -0.31169 &  & 0.00551 & -0.06925 & 0.01886 & & & & & \\ \hline
  \hline
\end{tabular}
\par
\end{center}
\noindent
{\bf Table S}2.
 Tight-binding parameters of the five-orbital model for
\text{RbFe$_{2}$As$_2$}. Here we
use the same notation as in Ref.~\onlinecite{SGraser:2010}.
The orbital index $\alpha=$1,2,3,4,5 correspond to $d_{xz}$, $d_{yz}$,
$d_{x^2-y^2}$, $d_{xy}$, and $d_{3z^2-r^2}$ orbitals, respectively.
The units of the parameters are eV.
\end{minipage}


\begin{minipage}{\linewidth}
\begin{center}
\begin{tabular}{cccccccccccc}
  \hline
  \hline
    & $\alpha=1$ & $\alpha=2$ & $\alpha=3$ & $\alpha=4$ & $\alpha=5$ &   & & & & &  \\ \hline
  $\epsilon_\alpha$ & 0.28734 & 0.28734 & -0.16395 & 0.62596 & 0.15918 &  & \\ \hline\hline
  $t^{\alpha\alpha}_\mu$ & $\mu=x$ & $\mu=y$ & $\mu=xy$ & $\mu=xx$ & $\mu=xxy$ & $\mu=xyy$ & $\mu=xxyy$ & $\mu=z$ & $\mu=xz$ & $\mu=xxz$ & $\mu=xyz$ \\ \hline
  $\alpha=1$ & -0.11178 & -0.27610 & 0.06091 & 0.23489 & -0.02882 & 0.03720 & 0.02428 &  & -0.00318 & 0.00724 & \\ \hline
  $\alpha=3$ & 0.35911 &  & -0.00843 & -0.08560 &  &  & & & & &  \\ \hline
  $\alpha=4$ & 0.09760 &  & -0.02986 & 0.09725 & 0.00281 &  & -0.07308 & 0.09810 & 0.04760 & & 0.01886 \\ \hline
  $\alpha=5$ & -0.04504 &  & -0.11908 &  & 0.01105 &  & 0.03144 & 0.02875 & -0.01499 & & \\ \hline\hline
  $t^{\alpha\beta}_\mu$ & $\mu=x$ & $\mu=xy$ & $\mu=xxy$ & $\mu=xxyy$ & $\mu=z$ & $\mu=xz$ & $\mu=xyz$ & $\mu=xxyz$ & & & \\ \hline
  $\alpha\beta=12$ &  & -0.09741 & -0.01353 & -0.03424 &  &  & -0.02276 & & & & \\ \hline
  $\alpha\beta=13$ & -0.45003 & 0.05198 & -0.03004 &  &  &  & & & & & \\ \hline
  $\alpha\beta=14$ & 0.25147 & -0.05763 & 0.04722 &  &  & -0.00347 & -0.00706 & -0.00007 & & & \\ \hline
  $\alpha\beta=15$ & -0.23510 & -0.10405 &  & -0.00000 &  &  & 0.00193 & & & & \\ \hline
  $\alpha\beta=24$ &  &  &  &  &  & 0.01628 &  & 0.01794 & & & \\ \hline
  $\alpha\beta=34$ &  &  & -0.04939 &  &  &  & & & & & \\ \hline
  $\alpha\beta=35$ & -0.23635 &  & 0.03186 &  &  &  & & & & & \\ \hline
  $\alpha\beta=45$ &  & -0.27431 &  & -0.00219 & 0.01117 & 0.02361 & & & & & \\ \hline
  \hline
\end{tabular}
\par
\end{center}
\noindent
{\bf Table S}3.
 Tight-binding parameters of the five-orbital model for
\text{CsFe$_{2}$As$_2$}. Here we
use the same notation as in Ref.~\onlinecite{SGraser:2010}. 
The orbital index $\alpha=$1,2,3,4,5 correspond to $d_{xz}$, $d_{yz}$,
$d_{x^2-y^2}$, $d_{xy}$, and $d_{3z^2-r^2}$ orbitals, respectively.
The units of the parameters are eV.
\end{minipage}

\end{widetext}

%